# Muon spin rotation studies of niobium for superconducting RF applications

A. Grassellino[1,2], C. Beard[1], P. Kolb[1,3], R. Laxdal[1], N.S. Lockyer[1], D. Longuevergne[1], and J. E. Sonier[4,5]

[1]TRIUMF, Vancouver, British Columbia, V6T2A3, Canada
[2]University of Pennsylvania, Philadelphia, Pennsylvania, 19104, USA[1]
[3]University of British Columbia, Vancouver, British Columbia, V6T1Z4, Canada
[4]Department of Physics, Simon Fraser University, Burnaby, British Columbia V5A 1S6, Canada
[5]Canadian Institute for Advanced Research, Toronto, Ontario M5G 1Z8, Canada

**Abstract**
In this work we investigate superconducting properties of niobium samples via application of the muon spin rotation/relaxation (µSR) technique. We employ for the first time the µSR technique to study samples that are cutout from large and small grain 1.5 GHz radio frequency (RF) single cell niobium cavities. The RF test of these cavities was accompanied by full temperature mapping to characterize the RF losses in each of the samples. Results of the µSR measurements show that standard cavity surface treatments like mild baking and buffered chemical polishing (BCP) performed on the studied samples affect their surface pinning strength. We find an interesting correlation between high field RF losses and field dependence of the sample magnetic volume fraction measured via µSR. The µSR line width observed in ZF-µSR measurements matches the behavior of Nb samples doped with minute amounts of Ta or N impurities. An upper bound for the upper critical field $H_{c2}$ of these cutouts is found.

## I. Introduction

One of the outstanding scientific issues related to the performance of SRF cavities made of high-purity niobium (Nb) (having a residual resistivity ratio RRR exceeding 200) is the occurrence of field-dependent RF losses in the cavity walls. Fig. 1 shows a typical plot of the quality factor Q of a 1.3-1.5 GHz Nb cavity versus the peak magnetic field at the resonator surface. There are three characteristic regions of RF losses in the absence of field emission, which are commonly referred to as: (i) the low-field Q slope (LFQS) occurring below 20 mT, (ii) the medium-field Q slope (MFQS), observed between 20 and ~ 80 mT, and (iii) the high-field Q slope (HFQS), which is observed above 80-100 mT. The RF losses typically increase at a gradual rate in the peak surface magnetic field range 20-80 mT, and increase sharply above ~ 80-100 mT. Thermometry tests of SRF Nb cavities [1, 2] show that high field losses (HFQS) always appear in the high magnetic peak field region. This indicates that it is the magnetic field component that is responsible for the losses in the Nb cavity walls in the HFQS regime.

Over the past decade, several models have been proposed to explain the field dependent RF losses (see Ref. [3, 4] for a review of these). However, none of these theories can explain the totality of experimental data, or suggest a means of experimentally validating a particular mechanism. Understanding the root cause(s) of the RF losses is important for improving the performance of SRF Nb cavities used in particle accelerators. Losses in the low and medium field regime determine the dynamic heat load to the cryogenic system and hence affect capital and operating costs for continuous wave accelerators like the proposed Next Generation Light Source, Project X, ERLs and others [5]. Losses at high fields determine the ultimate limitation for pulsed machines like the proposed International Linear Collider (ILC) where the operating gradient defines the accelerator length for a given final energy.

---
[1] Now at Fermi National Accelerator Laboratory, Batavia, IL, USA, annag@fnal.gov

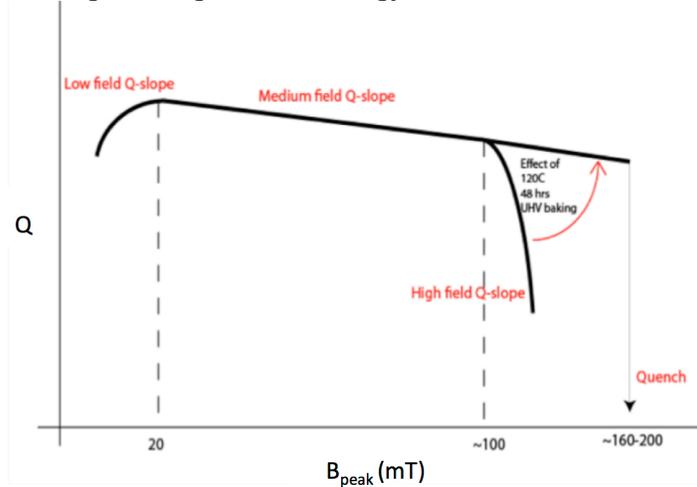

**Figure 1.** Typical Q-curve for a 1.3 or 1.5 GHz Nb resonator, showing the three magnetic-field regions where the magnetic field dependence of the quality factor changes: (i) the LFQS below 20 mT, (ii) the MFQS between 20 and 80-100 mT, and (iii) the HFQS above 80-100 mT. Also shown is the typical effect of a 120°C 48 hrs UHV bake on the HFQS.

To gain a better understanding of the origin of the RF field dependent losses, we study the superconducting properties of samples with identical structure to the cavities in question. The samples are cut out from large and small grain 1.5 GHz Nb cavities. The technique employed for conducting these studies is µSR (the acronym stands for muon spin rotation, relaxation and resonance), which involves implanting spin-polarized muons into a sample and following the time-evolution of the polarization as a precise diagnostic of the local magnetic field.

Both the tool and the samples used in this study are unique and offer a new and innovative approach: it is the first time that a measurement of flux penetration and flux evolution inside samples which have been cutout of SRF cavities is attempted. This is crucial since those samples have the exact crystalline structure/impurity content which cause RF losses, and moreover the RF losses have been well characterized via thermometry during the RF tests (as it will be described more in detail in the following paragraph). The technique brings the advantage that the probe is planted inside the sample and can provide therefore unique information e.g. the sample magnetic volume fraction as function of magnetic field surrounding the sample.

Some of the samples were measured before and after being subjected to typical treatments known to affect RF losses (such as 120 °C UHV baking and buffered chemical polishing, BCP). Specifically, a 120 °C bake in UHV for 48 hours is known to often eliminate or mitigate high-field RF losses, by pushing the onset of the HFQS to higher peak surface magnetic field [6], and also to increase the residual component and decrease the BCS component of the RF surface resistance. The effect of the 120 °C bake on high field Q-slope is also known to be confined to the first tens of nm from the surface, and in the first ~ hundred for BCS surface resistance [7, 8, 9], so a procedure like BCP completely reverses the benefit of the bake. By repeating the µSR measurements before and after such treatments for the same samples, we measure the relative change in magnetic field penetration and magnetic volume fraction due exclusively to the effect of surface treatments and free from any geometrical effect.

The layout of the paper is as follows. In Sec. II we briefly describe the RF characterization of the samples via thermometry measurements performed at Cornell University [1] and the μSR experimental setup. The μSR measurements results are described in Sec. III followed by discussion in sec. IV.

## II Experimental Details

### A. Thermometry characterization of RF losses

Tests of superconducting Nb cavities to determine the dependence of the quality factor Q on the strength of the applied RF field can be performed with a temperature mapping system attached to the outer cavity walls. This enables a determination of both the location of the RF losses and the RF field level at which the losses occur. Thermometry studies are essential for understanding and distinguishing the source of the RF losses; whether for example they stem from field emission, multipacting or Q-disease can be revealed by characteristic heating patterns [10]. Several studies have been done to understand how the typical RF losses are spatially distributed. These studies [2, 11, 12] indicate that in the HFQS regime the heating pattern at the surface always appears in the high magnetic field regions, and is spatially non-uniform, with some patches hotter than others (Fig. 2). Several tests are ongoing for studying the heating pattern in the medium field regime and some first results indicate that some losses at medium field might have the same origin of the high field Q-slope losses [5], and therefore also originate from surface magnetic peak fields.

Therefore, RF tests coupled with thermometry measurements can identify the regions of the cavities that exhibit higher or lower losses as a function of the RF field. Once characterized, these regions can be cut out of the cavity for subsequent investigation. Several investigations have been done on so-called 'hot' (higher losses) versus 'cold' (lower losses) samples, and baked versus unbaked HFQS-limited cutout samples, to examine differences in the microscopic structure. Some of the techniques employed include electron back-scattered diffraction and positron annihilation spectroscopy [13]. The results of some of these sample studies [14, 15] show a higher density of lattice defects in unbaked compared to baked samples. This suggests that lattice defects play an important role in the cavity RF losses.

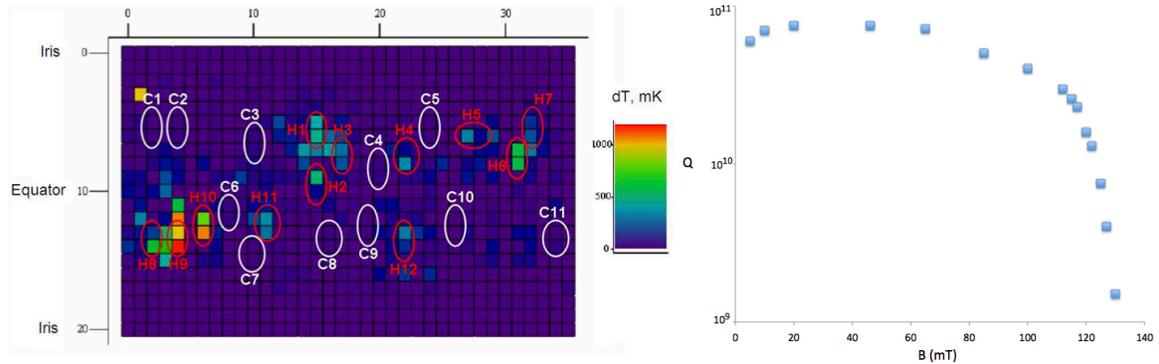

**Figure 2.** Thermometry map and RF test of a 1.5 GHz large grain Cornell Nb cavity LE1-37. Temperature map recorded at a peak surface magnetic field of 130 mT, which is in the HFQS regime. Losses range from a few mK in the 'cold' spots to several hundreds of mK in the 'hot' spots. Courtesy of A. Romanenko [1].

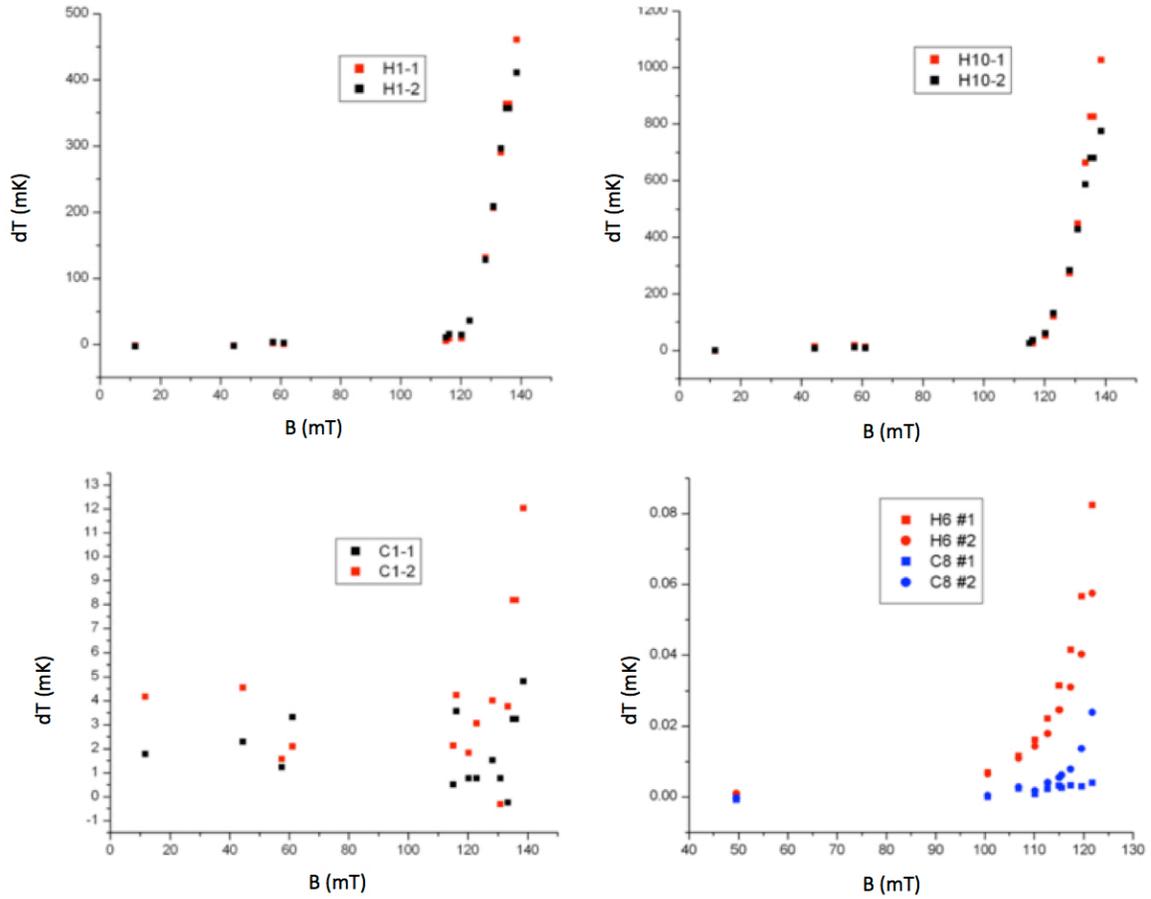

**Figure 3.** Thermometry characterization of RF heating for the samples studied (H1, H6, H10, C1). Courtesy of A. Romanenko [1].

The samples studied here by μSR were cut from two 1.5 GHz Nb cavities, (large grain and small [1 mm] grain), which have undergone buffered chemical processing (BCP) of the surface. Circular samples of diameter ~20 mm and thickness 2.8 mm were cut out from regions of both elevated and reduced heating. Details on the dissection procedure are described in Ref. [1]. The Cornell temperature mapping system attached to the outer walls of the intact cavities during RF testing consists of an array of 756 thermometers. Each thermometer measures the temperature difference between a spot on the outer surface of the cavity and the helium bath. A detailed description of the thermometry system can be found in [1]. The samples are characterized by their temperature as a function of the RF magnetic field. Fig. 3 shows heating map results for five different samples — including three cutouts from the large grain BCP cavity (H1, H10, C1), and two samples (H6, C8) that were cutout of a small grain (1 mm grain) BCP cavity.

**B. μSR setup**

The μSR measurements were carried out on the M20 surface muon beamline at TRIUMF using the LAMPF spectrometer, which consists of three pairs of Helmholtz coils and a helium-gas flow cryostat. The μSR technique applied in our studies is a sensitive probe of local internal magnetic fields that utilizes a beam of short-lived spin-polarized positive muons. With kinetic energies near 4.1 MeV (momentum 28 MeV/c), the spin-polarized surface muons stop in the bulk of the sample. TRIM simulations show that muons with this momentum have a mean stopping depth in pure Nb of ~ 300 µm (see Fig. 4). The muon beam has a typically Gaussian width σ ~ 5mm so does not allow fine positional resolution although Ag masks can be used to localize the implantation area. Coexisting magnetic and non-magnetic regions in the same specimen are recognizable by distinct components of the µSR signal whose amplitudes are proportional to the volume of the sample occupied by each phase. Thus µSR provides quantitative information on the magnetic volume fraction.

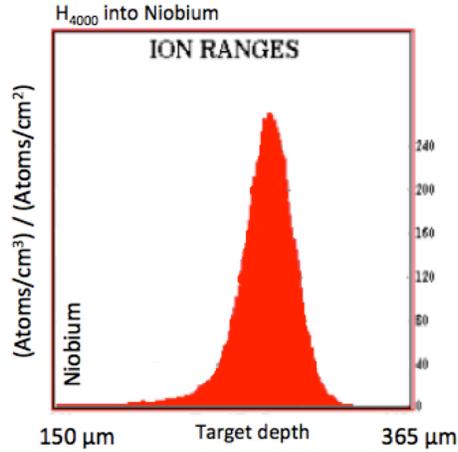

**Figure 4.** TRIM simulation of the muon stopping profile in pure Nb, showing a peak at around 300 microns from the surface.

The implanted muons stop at interstitial sites in the Nb crystal lattice where they Larmor precess at a frequency proportional to the magnitude of the average local magnetic field ($\omega = \gamma_\mu B$). The time evolution of the muon spin polarization $P(t)$, which is recorded via the detection of the decay positrons, reflects the internal magnetic field distribution. The measurements were carried out under field-cooled or zero-field cooled conditions in either a "transverse field" (TF-µSR) or "zero field" (ZF-µSR) geometry. In the TF-µSR experiments, a static magnetic field was applied along the beam direction ($z$ axis), perpendicular to the initial muon spin polarization **P**(0) (see Fig. 5) and perpendicular to the sample. The ZF-µSR experiments were instead performed with **P**(0) parallel to the beam direction using "front" and "back" positron detectors (not shown in Fig. 5).

Since the external magnetic field lines are forced to bend around the edges of the sample, the critical field for magnetic flux penetration near the sample perimeter is reached at lower values of the applied magnetic field. With increasing magnetic field we expect the magnetic flux line segments penetrating the edges to eventually be driven to the center by the surface supercurrents. This process will be strongly affected by the surface pinning. Therefore, our measurements will carry information on the pinning strength at the surface of each sample as a function of different surface treatments. In most of our reported results only the central region of the sample was exposed to the muon beam. This was achieved by sandwiching a pure silver (Ag) mask, having a circular center hole of 8 mm in diameter, between two incoming muon detectors, as shown in Fig. 5. This way, muons are prevented from stopping in regions of the sample within 6 mm of the sample edges. Muons passing through the first muon counter, but not the second, stop in the Ag mask, and hence are logically excluded via the data acquisition electronics. The Ag mask is required to ensure that the implanted muons detect the field in the center of the sample that is independent from sample edge effects. For one sample (H1) the positional dependence of muon application was studied by repeating the tests with no mask so the muons position was not restricted. Another sample C1 was measured with the central mask and an annular mask, a ring from 4 to 6mm radii. As it will be shown in the next paragraph, a comparison of the measurements with and without the silver mask, and with the central versus annular mask, gives information on the mechanisms and field thresholds for flux entry and flux evolution in the samples.

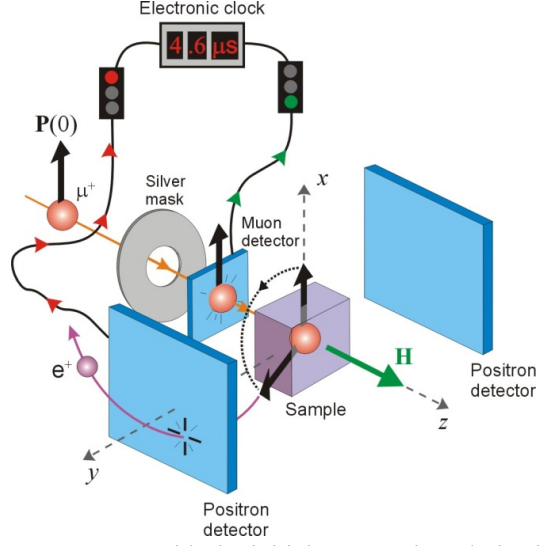

**Figure 5.** Schematic of the TFµSR setup with the initial muon spin polarization **P**(0) perpendicular to the applied static magnetic field **H**. The silver mask positioned upstream ensures that the positive muons (µ$^+$) stop only in the central region of the sample.

The TF-µSR geometry was used to study magnetic-flux penetration at temperatures of $T$ = 2.3 K, and $T$ = 4.2 K to 8 K. Each measurement was done by implanting approximately $2 \times 10^7$ spin-polarized muons one at a time into the sample. The µSR signal is given by

$$A(t) = a_0 P_i(t) \qquad (1)$$

where $A(t)$ is called the µSR "asymmetry" spectrum, $a_0$ is the initial asymmetry, and $P_i(t)$ is the time evolution of the muon spin polarization. The subscript for the TF-µSR and ZF-µSR arrangements is $i = x$ and $i = z$, respectively. For example, in the TF-µSR configuration

$$P_x(t) = \int_0^\infty n(B)\cos(\gamma_\mu B t + \phi)dB \qquad (2)$$

where $\phi$ is a phase constant and

$$n(B') = \langle \delta(B' - B(r)) \rangle \qquad (3)$$

is the probability of finding a local field $B$ in the $z$ direction at an arbitrary position $r$ in the $x$-$y$ plane.

## III. Results

### A. ZF-μSR measurements

Figure 6(a) shows the ZF-μSR signal from sample H1 at several different temperatures. In zero external magnetic field, the ZF-μSR signal is expected to relax by the sensitivity of the muon to the randomly oriented nuclear moments. Because the correlation times of the nuclear moments are much longer than the muon lifetime, they are generally seen as a dense static Gaussian-like internal field distribution that does not change with temperature. However, this is clearly not the case in Fig. 6(a). Instead the ZF-μSR signals are best described by a *dynamic* theoretical relaxation function. In particular, the solid curves through the data in Fig. 6(a) are fits to a *dynamic* Gaussian Kubo-Toyabe function [16], which is characterized by a hop rate $v$ associated with the variation in local magnetic field experienced by a mobile muon. The temperature dependence of $v$ is shown in Fig. 6(b) for the H1 sample. No significant difference is observed between the data collected before and after baking. A similar temperature dependence of the μSR line width has been observed in ZF-μSR and TF-μSR measurements on Nb samples doped with minute amounts of Ta or N impurities [17, 18]. (Note, that the data in these earlier works was fit less accurately with $v = 0$ and the μSR line width was free to vary with temperature). The behavior can be explained by a model of muon diffusion whereby the positive muon thermally breaks free from an extended shallow trap site above $T \sim 20$ K, quickly diffuses to a deeper more localized trap site at $T \sim 50$ K, but escapes this second trap site and becomes increasingly more mobile at higher temperatures. Positive muons are expected to react like a hydrogen ion p+ and they might be sensitive to the same chemical interactions. From extensive internal friction studies around these temperatures, hydrogen is known to interact respectively with dislocations and/or to exhibit specific H-H, H-O, or H-N interactions [19, 20, 21]. Repeating the same studies at the surface of the cutout samples with low energy μSR, where muons stop in the first tens of nanometers, could provide important insights on the role of hydrogen in RF losses.

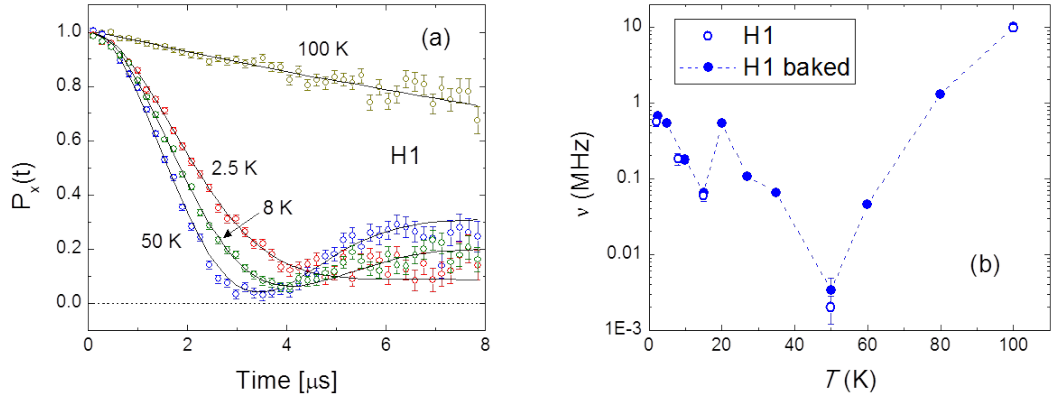

**Figure 6.** (a) Representative ZF-μSR spectra of sample H1 at different temperatures. (b) Temperature dependence of the muon hop rate in sample H1 before and after baking.

### B. TF-μSR measurements

TF-μSR measurements were taken on all samples after zero field-cooling at a temperature of 2.3 K and subsequently applying a static magnetic field in the range 5 mT to 270 mT. Two of the samples (H1 and C1) were subsequently baked for 48 hours at 120°C in UHV and re-measured by μSR. Measurements were also performed on the baked H1 sample subsequently processed with BCP, which leads to the removal of a 5 micron thick layer at the surface. Fig. 7 shows representative TF-μSR signals at ~ 30 mT and 120 mT. At 30 mT the TF-μSR signals for each sample closely resemble the ZF-μSR signal, indicating that the applied magnetic field is completely screened from the bulk of the central region of the sample. At ~ 120 mT the amplitude of the TF-μSR signal, which is a measure of the magnetic flux-free volume fraction, is reduced. Note that the fraction of the muons that sense magnetic flux in the sample actually contribute to a rapidly relaxing component, which can be observed at early times with significantly smaller time binning than that used to generate Fig. 7. Comparing Figs. 7(b) and 7(c), we see that the volume fraction of the sample penetrated by the external magnetic field is reduced after the 120 °C baking. However, as shown in Fig. 7(d), this is reversed by subsequent BCP of the surface. Since the TF-μSR signals displayed in Fig. 7 correspond only to those muons that do not sense magnetic flux in the sample, they are also well described by a dynamic Gaussian zero-field Kubo-Toyabe relaxation function.

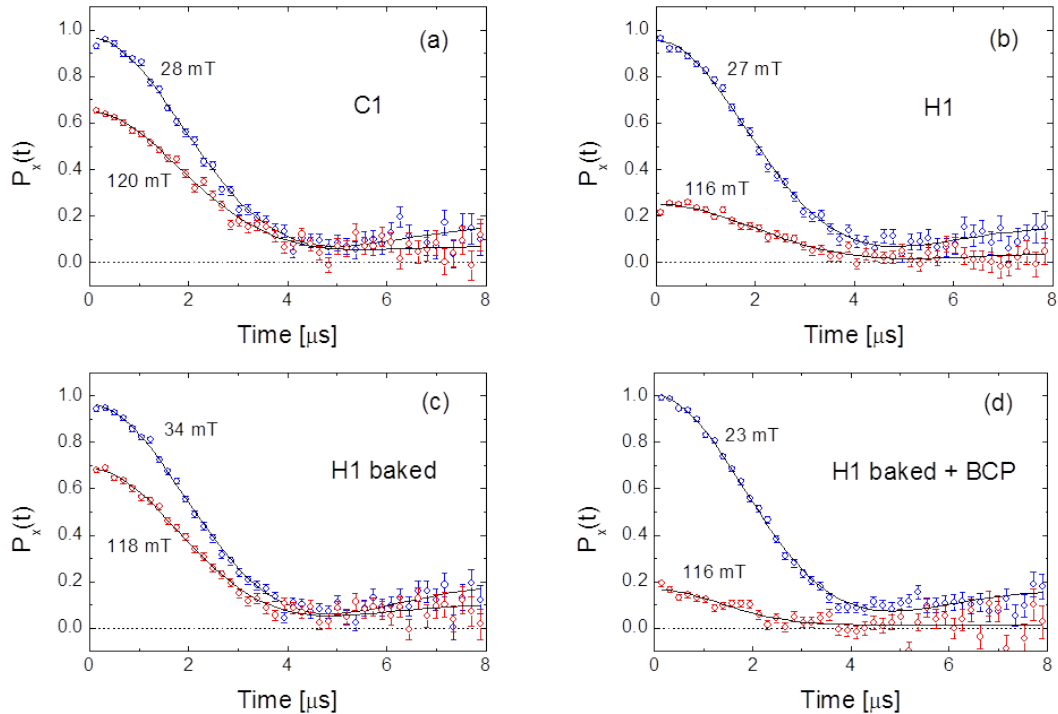

**Figure 7.** TF-μSR spectra recorded at $T$ = 2.3 K and at applied magnetic fields of approximately 30 mT and 120 mT, for samples (a) C1, (b) H1, (c) H1 after 48 hours of UHV 120 °C baking, and (d) H1 after baking followed by BCP.

Figure 8(a) shows the corresponding fitted amplitude $a_0$ associated with the volume fraction of magnetic-flux free regions, as a function of applied magnetic field for the H1 sample - as cut from the large-grain cavity and after being subject to 120°C baking and BCP. The characteristic Q-curve of the cavity from where the sample was cut is shown in Figure 8(b). It is interesting to note the correlation between the onset of the HFQS and of the increase in fraction of magnetic flux in the central region of the sample measured by μSR, both around 100 mT. Fig. 9 shows the corresponding fitted amplitude $a_0$ as a function of applied magnetic field for all the samples measured. All the samples, hot and cold, show an increase in magnetic volume fraction at 2.3K at around 100 mT external field (applied perpendicular to the surface). In the cold sample the increase in magnetic volume fraction has same onset but grows slower than in the hot sample, as shown in fig. 10. H1 and C1 were re-measured after 120°C bake, and show that the increase in magnetic volume fraction at 2.3 K shifted to ~ 125 mT (fig.10), identically for hot and cold sample. Finally, sample H1 underwent a further processing step of 5 micron surface removal via BCP. The increase in magnetic volume fraction occurred then at ~ 85 mT, a 40 mT change compared to the previous 125 mT post 120°C bake.

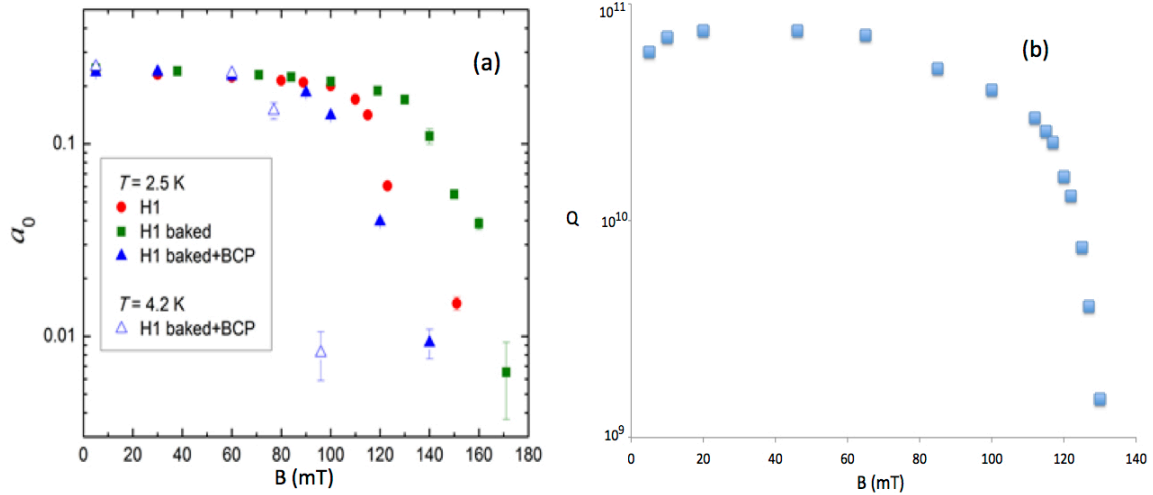

**Figure 8.** (a) Initial asymmetry $a_0$ of sample H1 associated with the volume fraction of magnetic-flux free regions. (b) Original RF characterization of the cavity.

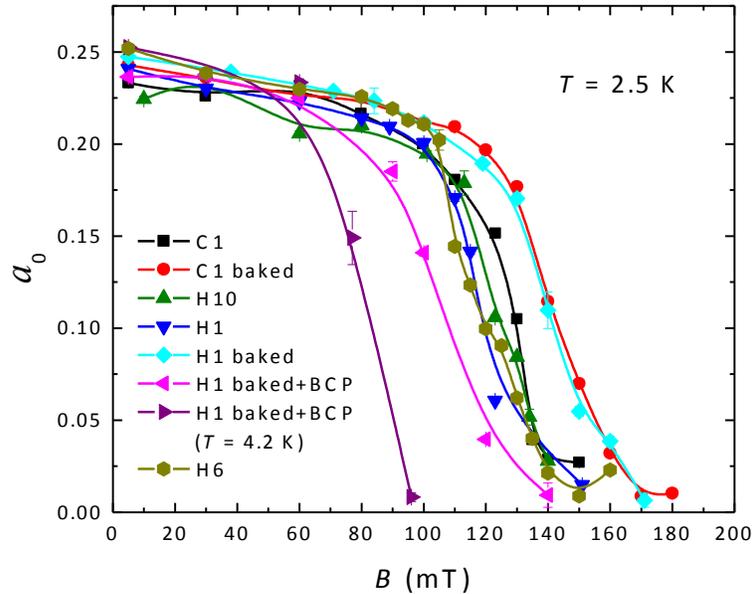

**Figure 9.** Magnetic field dependence of the volume fraction of the sample not containing magnetic flux (which is proportional to $a_0$). Results are shown for all the samples tested at $T \sim 2.3$ K and for the H1 sample at $T = 4.5$ K after baking and BCP (purple symbols).

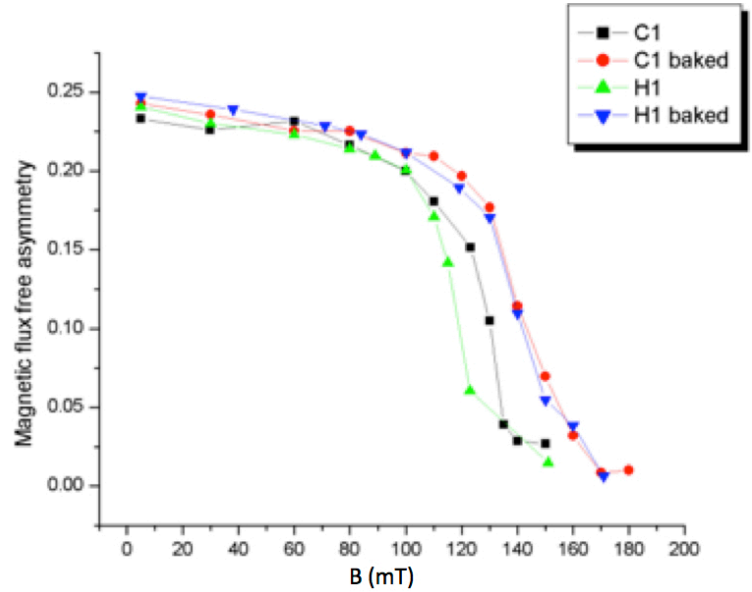

**Figure 10.** Identical to Fig. 9 but highlighting the performance of samples H1 (hot spot cutout from the large grain cavity) and C1 (cold spot cut from the same cavity) measured at 2.3 K, before and after 120°C baking.

All the measurements reported so far are for muons implanted in the central region of the sample, via a silver mask with a central hole of radius 4 mm. To gain a better understanding of how the flux first enters and how it evolves in the sample, we repeated the TF-μSR measurements on the same sample H1 baked plus BCP without any mask (so muons are implanted in the whole sample) and with an annular mask (a concentric mask letting muons implant through a ring going from 4 to 6 mm radius).

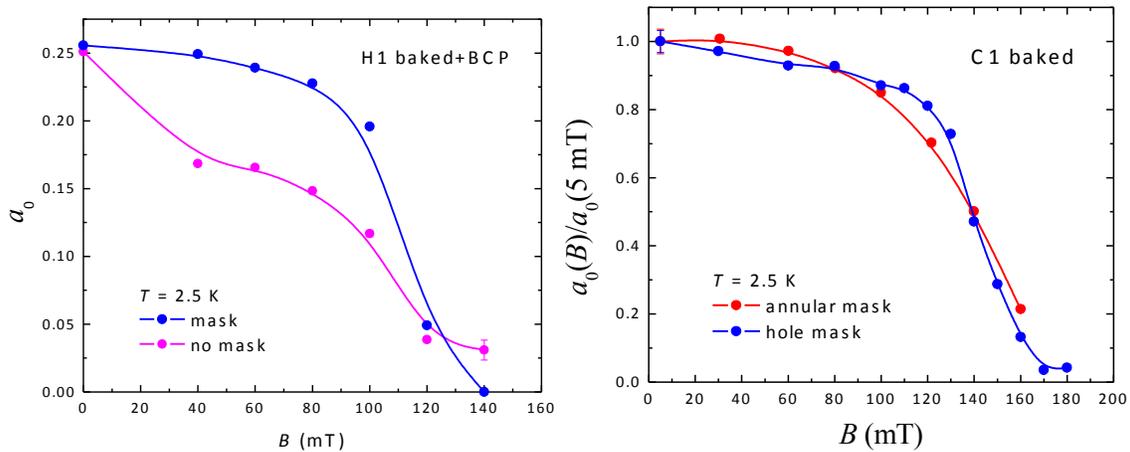

**Figure 11.** Left: initial asymmetry $a_0$ of sample H1 baked plus BCP associated with the volume fraction of magnetic-flux free regions, comparison of the measurement with the central silver mask and without the silver mask. Right: normalized asymmetry of sample C1 baked, comparison of measurements with two different masks, central (hole) and annular.

The left plot on Fig.11 shows the comparison of the magnetic volume fraction measurements for the same sample H1 baked plus BCP for the central mask and no mask configurations. This shows how the magnetic volume fraction in the case with no silver mask increases below 40 mT, showing that flux indeed penetrates the samples edges early. Then, above 40 mT there is no increase in magnetic volume fraction up to the same threshold field point as the measurement with the central mask, which is around 100 mT. The right plot in Fig. 11 shows the comparison of the measurements taken for sample C1 baked with the central mask and with an annular mask, a ring from 4 to 6 mm radii. Also in this case, it is clear the sharp increase in magnetic volume fraction occurs at around the same field 100 mT for both masks (central and ring). In the next section we will discuss these results in detail, along with possible interpretations.

The internal magnetic field distribution can be visualized in a fast Fourier transform (FFT) of the μSR asymmetry signal, which is given by

$$n(B) = \int_0^\infty P(t)e^{-i(\gamma_\mu Bt+\theta)}dt . \qquad (4)$$

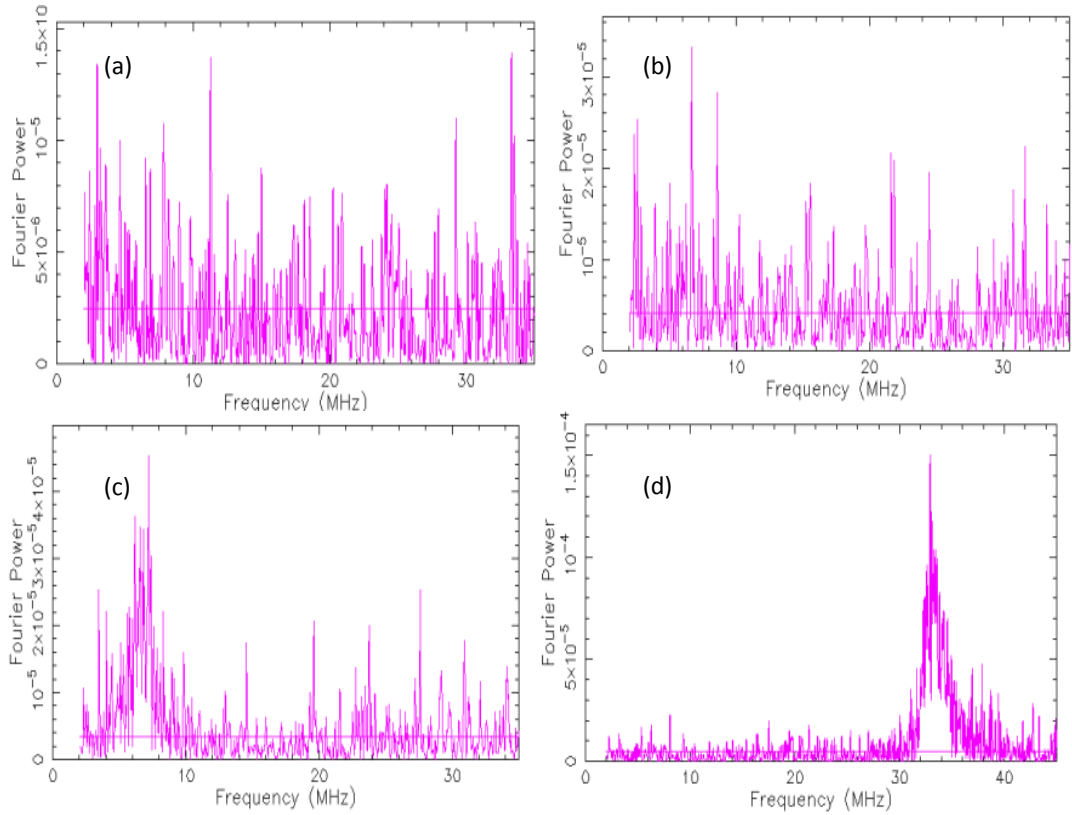

**Figure 12.** Fast Fourier transforms n(B) vs muons precession frequency of the μSR signal from sample H1 at $T = 2.3$ K and at an applied field of (a) 0 mT, (b) 30mT, (c) 120 mT, and (d) 270 mT. The frequency is proportional to the internal magnetic field sensed by the muon.

The field value corresponding to a particular frequency in the FFT spectrum is found via the muon gyromagnetic ratio and is given by B(mT) ~ 7.3*f(MHz). Fig. 12 shows the FFTs of the μSR signal from sample H1 at $T = 2.3$ K and for several different values of the applied magnetic field. At 120 mT there is a peak appearing in the FFT at ~ 7 MHz frequency, corresponding to an average internal magnetic field of approximately 50 mT. This peak shifts upwards with an increase in the magnitude of the applied magnetic field, and this is accompanied by a reduction in the difference between the external and internal fields. At an applied field of 270 mT, the peak in the FFT is located at a frequency of 33 MHz corresponding to ~ 240 mT.

**Measurement of upper critical field Hc2**

TF-muSR measurements were completed on sample H10 at temperatures T= 2.3 K, 4.2 K and T=7.5 K for different field levels to localize the transition to full normal conducting state. This transition should occur at the upper critical field $H_{c2}$. As shown in eq. (4), the Fourier transform of the asymmetry signal, often called the µSR line shape, shows the internal magnetic field distribution for a certain applied external field, where the Fourier power is proportional to the fraction of muons sensing a certain magnetic field inside the sample. In the vortex state, the muons sense a field 'range' which goes from the value of the magnetic field inside a vortex (in a region of the size ~ coherence length ξ) and which decays to zero exponentially over a length λ. In the normal state, the muons sense only one field, which, since the sample is normal conducting, corresponds to the external applied magnetic field, hence the FFT will be a sharp line at the applied field. A description of typical µSR line shapes can be found in [22].

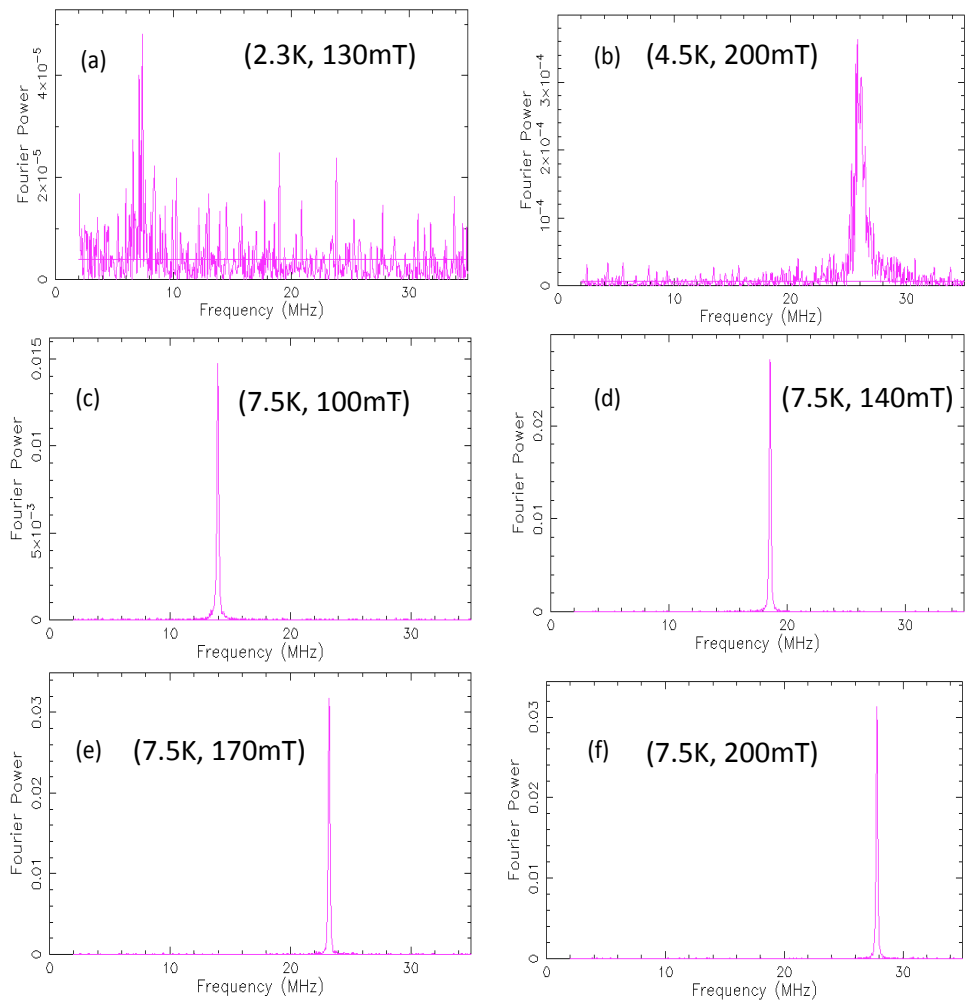

**Fig.13** FFTs for sample H10 respectively for temperature and fields: a (2.3K, 130mT), b (4.5K, 200mT), c (7.5K, 100mT), d (7.5K, 140mT), e (7.5K, 170mT), f (7.5K, 200mT).

Fig.13 shows in order the FFT for the H10 sample asymmetry signal for 130 mT at 2.3 K, 200 mT at 4.5 K and 100 mT, 140 mT, 170 mT, 200 mT at 7.5 K. The first plot (2.3 K, 130 mT) shows the presence of magnetic flux in the sample, with peaks appearing at different field levels, indicating magnetic flux presence in the sample but not in the form of a homogenous ordered vortex lattice. The second plot (4.5 K, 200 mT) shows that the sample is now deep in the vortex state, with the peak of FFT appearing at the ~ applied field of 200 mT. At (7.5 K, 100 mT) the sample is turning completely normal conducting, although in the following points at same T but higher field, the amplitude of the peak keeps growing saturating to full amplitude at (7.5 K, 170 mT). An analytical expression for the upper critical field as a function of temperature for a type-II superconductor is [23]:

$$H_{c2}(T) = H_{c2}(0) \cdot \frac{1-(T/T_c)^2}{1+(T/T_c)^2} \quad (5)$$

If we choose (7.5 K, 100 mT) as the transition point to normal conducting state this would correspond to the point (4.5 K, 388 mT) based on (5). As shown in the second plot of Fig.12 at (4.5 K, 200 mT) the sample is in the vortex state, so it's in between $H_{c1}$ and $H_{c2}$, which sets a lower bound for $H_{c2}(0)$= 325 mT. Choosing $H_{c2}$(7.5 K) = 100 mT we establish an upper bound for $H_{c2}$, which based on eq. 5 corresponds to $H_{c2}$(4.5 K) = 388 mT, $H_{c2}$(2.3 K) = 466 mT and $H_{c2}$(0) = 496 mT. The data might suggest a scenario in which regions of different critical fields $H_{c2}$ coexist in the sample, ranging up to $H_{c2}$(7.5 K)=170mT, point at which the whole sample has turned normal conducting.

## IV. Discussion

We investigated, for the first time using the muon spin rotation technique, the superconducting properties of niobium samples, which have been cutout of 1.5 GHz niobium RF cavities, whose surface was treated via buffered chemical polishing. Several magnetization measurements exist [24] on RRR 300 niobium samples, subject to different typical cavities surface treatments. However, samples cut out of a Nb sheet do not have the exact crystalline structure/impurity content which uniquely determines the superconducting behavior, and therefore the RF losses, of SRF cavities (a summary of the processes from raw sheet to formed cavity can be found in [3]); also, the behavior of samples from niobium sheets under a RF field is unknown, and therefore all the existing studies are under an unproven assumption that these samples would have the same RF losses as SRF cavities. The measurements presented in this paper add therefore new information to the existing measurements on RRR 300 sheet samples, being the samples used cutouts from a real cavity after RF test and thermometry characterization. Moreover, the technique uses muons implanted 'inside' the sample as an extremely sensitive local probe of magnetic fields. Muons can detect the magnetic volume fraction of the sample as function of applied field and hence can carry information on how magnetic flux first enters and then evolves in the studied samples as a function of the applied external magnetic field.

We first study the sample under no applied field, via ZF-μSR. When no field is applied to the sample, the signal is expected to relax by the sensitivity of the muon to the randomly oriented nuclear moments. The ZF-μSR measurements carry information about bulk properties, since the muons stop at around 300 microns from the surface and probe the local fields environment at that depth. The asymmetries spectra resemble that of Nb samples doped with minute amounts of Ta or N impurities [17, 18], which is consistent with niobium used for SRF cavities, which typically is of RRR ~ 300. There is no difference between the spectra recorded for the same sample before and after the 48 hours UHV 120 °C bake, which also is consistent with the fact that the 120 °C bake affects the SC properties of the surface (first hundreds of nanometers)[7] but not the bulk.

We then investigate the magnetic volume fraction of the samples in the TF-µSR configuration, as a function of the applied magnetic field perpendicular to the sample surface. In this case, even though the muons are implanted ~300 microns from the surface, the experiment can provide information on changes occurring at the surface, since the screening currents, which determine when flux enters and reaches the muons, will be flowing in the London layer (first tens of nanometers). When the applied field $B_a$ increases from zero, the sample initially is in the Meissner state, and no magnetic flux penetrates except to a depth of lambda, the London depth. The magnetic field at the edge of the sample is larger than $B_a$ due to demagnetization effects. Flux (vortex) penetration starts when the maximum field at the surface reaches $B_{c1}$. At that moment the applied field is smaller than $B_{c1}$ due to demagnetization, namely $B_{en}$. The field at which magnetic flux penetrates $B_{en}$ was found in [25] for a disk shaped sample of thickness b and diameter a to be:

$$B_{en} = B_{c1} \cdot \tanh \sqrt{0.67 \cdot \frac{b}{a}} \qquad (6)$$

Substituting in equation 6 the typical parameters for our disk shaped samples of b ~ 0.3 cm and a ~ 2 cm, and assuming a typical $B_{c1}(2K)$ = 170 mT for RRR ~ 300 niobium [26], we obtain that $B_{en}$ = 50 mT. So we expect that already at applied field of 50 mT flux penetrates the edges of the sample. This is indeed confirmed by the measurement of the sample H1 baked plus BCP where no silver mask is present (and therefore muons probe magnetic flux throughout the sample, including the edges). As it can be seen in the left plot of Fig.11, the magnetic volume fraction actually increases in the measurement with no mask already between zero and 40 mT applied field. This might indicate that $B_{c1}(2K)$ is actually even lower than 170 mT.

Before $B_{en}$ is reached, the bulk is free of vortices (B=0). Once $B_{en}$ is reached, in the corners of the sample there "hang" vortices. These will be drawn towards the disk center by the surface screening current. This force is balanced by the elastic tension of the vortex that draws them outwards. In a pin free case, flux will jump to the center of the sample, from where it gradually fills the disk, as described in [25]. If there is weak pinning though, this jumping to the center is delayed, and the viscosity determines the velocity of the moving vortices [27]. The scenario of sudden jump to the center or to a pinning-caused delayed flux motion is well established in numerous experiments and calculations [28, 29, 30]. In our TF-µSR studies, we performed measurements on same samples (so no change in geometry) before and after typical cavity surface treatments, therefore a comparison of the onset of appearance of flux in the center region of the sample provides information on how the surface pinning is changing with typical cavity treatments. A comparison among different samples is also valid, since all samples are geometrically almost identical.

To better understand the flux entry and evolution in the cutout samples studied, we repeated the measurements for two samples with no mask or with an annular mask, as described in the previous paragraphs. As we can see from Fig.11, in sample H1 baked plus BCP, magnetic flux appears at the edges below 40 mT applied field. Then, all H1 measurements with the central silver mask and without it, show no significant increase in magnetic volume fraction up to the same threshold of $B_a$ ~ 100 mT. This indicates that the sample pinning is quite weak, and it is consistent with a scenario where at $B_{en}$ (below 40 mT) flux enters the edges and then 'hangs' there until it can overcome the weak pinning force, at a field level that we name $B_{pin}$. Above $B_{pin}$ flux jumps to the center of the sample. Also, we notice that the magnetic volume fraction above $B_{pin}$ increases faster in the central region than in the whole sample, indicating that the flux jumps to the center and fills the sample from there outwards. Same can be concluded from the measurements performed on the baked sample C1, with different masks. As it can be seen in the right plot of fig. 11, both central mask and annular mask measurements show that the fast increase in magnetic volume fraction occurs at the same field level ~ 100 mT, indicating again that the flux front does not move progressively from the edges to the center, but rather suddenly jumps above $B_{pin}$ from the edges to the center, filling then the sample from the center outward. A similar behavior has been observed in a decoration experiment by Essmann [29], on a polycrystalline Nb sample of similar dimensions to that of our cutout samples. Flux first penetrates at the edges in form of fingers, and once passed the edge barrier jumps to the center and fills the sample from the center outwards.

Table 1 summarizes approximate values of $B_{pin}$ for the different samples measured, before and after certain typical cavity treatments, chosen as the onset points of the sharp increase in magnetic volume fraction. As we can see in the table, all samples as cut from the cavity exhibit very similar value of $B_{pin}$. The first thing we notice is the striking correlation between $B_{pin}$ measured in all samples and the onset of high RF field losses (HFQS), both at B ~ 100 mT, as shown in Fig. 8.

Table 1: Summary of approximate field levels at which magnetic volume fraction sharply increases in the central region of the sample. Last column indicates the sample (central region) volume percentage occupied by magnetic flux at 120 mT applied field.

| Sample | Type | $B_{pin}$ [mT] As cut, T=2.3K | $B_{pin}$ [mT] Post 120C bake, T=2.3K | $B_{pin}$ [mT] Post BCP, T=2.3K | Magnetic volume fraction at 120 mT |
|---|---|---|---|---|---|
| H1 | Large grain BCP, hot | 100 | 125 | 85 (T=2.3K) 60 (T=4.5K) | 60% as cut 20% post 120C 80% post BCP |
| C1 | Large grain BCP, cold | 100 | 125 | - | 35% as cut 20% post 120C |
| H10 | Large grain BCP, hot | 100 | - | - | 55% as cut |
| H6 | Fine grain BCP, hot | 105 | - | - | 60% as cut |

This strong quantitative correlation is at first surprising, considering that the two geometries are very different: the muon spin rotation measurement occurs on a disk shaped sample in perpendicular field (therefore with significant demagnetization effects) while cavity operation happens with the RF field applied parallel to the surface. One possible explanation for the correlation found despite the different geometry of the problem, is that since the cavity surface roughness ranges from nanoscale to micron scale -depends on the presence of grain boundaries and on surface treatments like electro-polishing (EP) or buffered chemical polishing (BCP)- there are unavoidable components of the applied RF field perpendicular to the cavity surface. Therefore, the µSR experiment may reproduce the same cavity surface situation but on a 'larger' scale, as conceptually illustrated in the fig.14. More generally, this correlation might indicate that the mechanism behind HFQS (and partially also MFQS) is linked to magnetic flux overcoming the typical cavity surface barrier. This has been partially conjectured already in the magnetic field enhancement model [31]. We will now put forward a qualitative model to explain losses in SRF cavities, based on the findings of this study, which goes beyond the magnetic field enhancement model by describing the surface barrier not only as a function of geometry, but as dictated by local lower critical field, local roughness and local pinning strength.

Losses in standardly treated SRF cavities typically follow two trends, in the medium field (20-80 mT) and high field range (above 80 mT) -the previously described MFQS and HFQS. We now put forward a qualitative model for medium and high field losses entirely based on dissipation due to magnetic vortices presence in the London layer. This model is based on two 'thresholds' for magnetic flux penetration, one at $B_{en}$, which sets the start of medium field losses, and one at $B_{pin}$, which corresponds to the onset of high field Q-slope. The model is conceptually illustrated in fig. 14. We suggest the possibility that already at lower fields (below the HFQS onset, but above the field of first entry $B_{en}$), some vortices can be present in the first few nanometers of the cavity surface. This is illustrated in fig.14 b. The μSR measurements might suggest that this flux starts being present in the cavity surface 'rougher' spots when the RF field reaches $B_{en}$, which, analogously to the muon spin rotation experiment, will be determined by the geometrical properties of the surface and by the local $B_{c1}$.

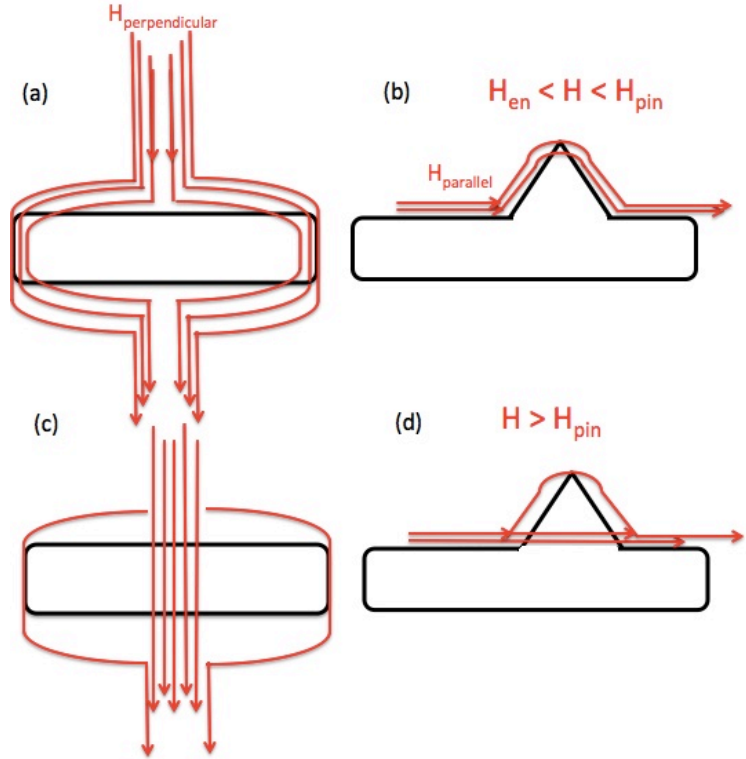

**Fig.14.** Schematic of a possible interpretation for the correlation found between HFQS onset in SRF cavities and the increase in magnetic volume fraction of the sample measured with μSR: a) μSR experiment, with magnetic field applied perpendicular to the sample; the magnetic field lines bend around (below $B_{en}$) and eventually penetrate the edges of the sample (above $B_{en}$); b) SRF cavity surface with magnetic field lines 'ideally' parallel to the surface, but with perpendicular components due to presence of surface roughness. Above $B_{en}$ some flux lines penetrate the first nanometers of rough spots, causing MFQS; c) above $B_{pin}$ magnetic field lines 'jump' to the middle of the sample and fill the sample from there outward; d) above HFQS onset, magnetic field lines, previously pinned at the rough edges, overcome the pinning force and jump inside the SC, causing HFQS.

So, the cavity surface will have some RF losses (MFQS) starting from lower RF fields, at the spots with lower $B_{en}$ as determined by:

$$B_{MFQS} = B_{en} = B_{c1\ local} \cdot f(g) \quad (7)$$

so proportional to the local lower critical field and a function of the local geometry of the surface. Similarly to how calculated in eq. 6 for the μSR case, the geometrical parameters that determine the surface geometrical barrier are the step height and angles, which determine the width of the step. Notice how geometry is not necessarily the only thing that determines a lower $B_{en}$, but also a lower local $B_{c1}$ can be responsible, like for example small islands of Nb-H precipitates, as described in [32]. As observed in the μSR experiment, this flux stays pinned in the first few nanometers, until it overcomes the pinning force as the applied RF field reaches $B_{pin}$ and starts being pushed in and out, causing high field Q-slope, as shown in fig. 14 d. The onset field of HFQS is then:

$$B_{HFQS} = B_{pin} = B_{c1\ local} \cdot f(g) \cdot f(p) \quad (8)$$

So the HFQS onset is determined by geometry, local critical field, and local pinning strength (p). This explanation for HFQS, like the above mentioned one based on magnetic field enhancement [31] calls roughness into play, and could be compatible with the many known experimental observations that HFQS onsets and cures do depend on the combination of grain size and chemical treatment (BCP gives a rougher surface than EP). But one point where the magnetic field enhancement model had failed in explaining the totality of experimental data is on why the 120°C bake removes the HFQS. The explanation proposed in this paper calls not only roughness but also pinning strength into play, which we will see from the experimental results, can also explain why the 120°C bake helps curing the HFQS.

There is a second possible scenario for explaining the correlation of our measurements and HFQS. A different possibility is that flux that is trapped during cooldown through $T_c$ (and that knowingly contributes to residual losses in SRF cavities) might stay pinned up to $H_{pin}$, after which it might start getting partially depinned and oscillating under the RF field, causing additional losses at high field (HFQS). This would imply that lossy locations at already low-medium field would be the same that would heat up in the HFQS regime. There is an indication that this might be true from some recent temperature mapping studies [5].

We proceed analyzing the results for the 120 °C bake, and discussing how they fit in the context of the two models just described and in general of cavity performance. Two samples (H1 and C1, hot and cold spot) were measured before and after the typical 48 hours UHV 120 °C bake. In both cases the effect of the 120 °C bake is an increase of about 25 mT in $H_{pin}$, to the exact same value for both samples. This result means that an effect of the 120 °C bake is to increase the surface pinning, which is in agreement with the hysteretic behavior of the magnetization curves of samples pre and post bake measured in [24]. The exact mechanism leading to the increase in pinning at the surface exceeds the scope of this paper, but might be consistent with current findings with positron annihilation experiments of an increase in near-surface vacancies post 120 °C bake [33]. An increase in surface pinning strength could be consistent with the observed increase in residual resistance in cavities post 120 °C bake, since increased pinning capability of the surface can lead to higher amount of trapped flux during cooldown through $T_c$, a known mechanism of residual losses in SRF cavities.

The increase in $H_{pin}$ post 120 °C bake found experimentally with this study can, most importantly,

explain the beneficial effect of the 120 °C bake on HFQS and is compatible with both previously described scenarios of flux entry or oscillating trapped flux above the $H_{pin}$ threshold. The 120 °C bake is known to increase the onset of HFQS in large grain BCP cavities, like the one where the studied samples were cutout from. The intuitive idea is that a surface with increased pinning strength will hold flux from 'breaking in' (fig. 14) or from 'moving' up to higher fields. This could be an important conclusion of this study, and suggestive of a potential way to increase achievable gradients in SRF cavities by 'engineering' the RF surface with enhanced pinning, for example by creating artificial pinning centers at the very near surface.

If we compare the results of the measurements for the hot sample H1 and the cold sample C1 as shown in fig. 10, we see that the onset of the increase in sample magnetic volume fraction is the same, but the fraction grows slower in the cold sample than in the hot one. As summarized in Table 1, at the applied field of 120 mT the hot sample H1 magnetic volume fraction is 60% versus 35% of the cold sample C1. This finding correlates very well with the RF losses characterization of sample H1 and C1 (fig.3), which shows how the onset of heating is the same for both samples, but the hot one heats up hundreds of mK and the cold one only few mK. This finding is also compatible with cavity performance and with both scenarios described above. In the first scenario (HFQS due to flux entry) a hot spot could have on average more sites that would cause premature flux entry, compared to a cold spot (for example average roughness higher for a hot sample than a cold one). Losses will onset when vortices at these rough sites will overcome the pinning force at the same field threshold $H_{pin}$, therefore onset of losses will be the same for hot and cold samples. But losses will be higher in the hot sample than in the cold one because of the higher amount of flux entering the surface in the first case than in the second. In the second scenario (HFQS due to trapped flux) hot spots could correspond to regions with (on average) more trapped flux than in the cold spots; again losses will onset at the same depinning field threshold $H_{pin}$ but will be higher in the hot one than in the cold one because of the higher amount of trapped flux in hot versus cold samples.

Moreover, H1 underwent a 5 microns removal via buffered chemical polishing, after the 120 °C bake. The measurement showed a decrease of 40 mT in $H_{pin}$, consistent with the fact that a light BCP indeed reverses the beneficial effect of the 120°C bake on high field losses. The measurement after BCP was also performed at 4.5 K. At this higher temperature, we observe that $H_{pin}$ is reduced to 60 mT, consistent with thermally activated depinning.

Finally, we estimated a lower and upper bound for for the upper critical field $H_{c2}$, which based on eq. 5 corresponds to $H_{c2low}(0) = 325$ mT and $H_{c2high}(0) = 496$ mT. The result is within the range of typically measured values of $H_{c2}$ for RRR 300 niobium [26].

## 5. Conclusions

We investigated, for the first time using the muon spin rotation technique, the superconducting properties of niobium samples, which have been cutout of 1.5 GHz niobium RF cavities. Results show a change in surface pinning with typical cavity treatments like mild baking and buffered chemical polishing. The findings correlate well with cavity performance pre and post treatments and possible mechanisms behind high and medium field losses have been suggested, based on depinning of trapped flux or breakdown of magnetic flux impinging at the cavity surface. These findings could have important implications on suggesting a possible pathway to increase achievable gradients in SRF cavities, based on artificially enhancing the surface pinning. An upper bound for the upper critical field of a cutout sample is found, which is in good agreement with previously measured values of for RRR 300 niobium. The muon spin rotation technique used for this study proved to be a proficuous experimental tool for investigating superconducting properties of niobium for SRF cavities.

**Acknowledgements**

The authors thank Dr. Alexander Romanenko (Fermilab) for providing the cutout samples, their RF characterization, and for invaluable discussions. We thank Dr Gerald Morris (TRIUMF), Dr Walter Hardy (University of British Columbia), Dr E.H. Brandt (Max-Planck) and Dr F. Barkov (FNAL) for several insightful discussions on interpretation of the results. We would like to thank Bassam Hitti and Donald Arseneau from the TRIUMF μSR group for their invaluable help during the experiment and Peter Harmer from the TRIUMF SRF group for the technical support with sample preparation. We thank Dr Camille Ginsburg and Dr Anthony C. Crawford from Fermilab for providing precious feedback on the manuscript.